# Model for Asteroid Regolith to Guide Simulant Development

Philip T. Metzger, Florida Space Institute, University of Central Florida, Orlando, Florida, USA, Philip.T.Metzger@ucf.edu

Daniel T. Britt, Department of Physics, University of Central Florida, Orlando, Florida, USA

**Abstract**

When creating asteroid regolith simulant, it is necessary to have a model of asteroid regolith to guide and to evaluate the simulant. We created a model through evaluation and synthesis of the available data sets including (1) the returned sample from Itokawa by the Hayabusa spacecraft, (2) imagery from the Hayabusa and NEAR spacecraft visiting Itokawa and Eros, respectively, (3) thermal infrared observations from asteroids, (4) the texture of meteorite regolith breccias, and (5) observations and modeling of the ejecta clouds from disrupted asteroids. Comparison of the Hayabusa returned sample with other data sets suggest the surficial material in the smooth regions of asteroids is dissimilar to the bulk regolith, probably due to removal of fines by photoionization and solar wind interaction or by preferential migration of mid-sized particles into the smooth terrain. We found deep challenges interpreting and applying the thermal infrared data so we were unable to use those observations in the model. Texture of regolith breccias do not agree with other data sets, suggesting the source regolith on Vesta was coarser than typical asteroid regolith. The observations of disrupted asteroids present a coherent picture of asteroid bulk regolith in collisional equilibrium, unlike lunar regolith, HED textures, and the Itokawa returned sample. The model we adopt consists of power laws for the bulk regolith in unspecified terrain (differential power index -3.5, representing equilibrium), and the surficial regolith in smooth terrain (differential power index -2.5, representing disequilibrium). Available data do not provide adequate constraints on maximum and minimum particle sizes for these power laws, so the model treats them as user-selectable parameters for the simulant.

**Introduction**

Regolith simulants are needed for a variety of engineering and scientific tests (Metzger et al., 2016). NASA requires that simulants be graded according to the Figure of Merit system (Schrader et al., 2009) as part of an attempt to stop the misuses that have occurred with lunar soil simulants (Taylor and Liu, 2010; Taylor et al., 2016). This is discussed further in "Status of Lunar Regolith Simulants and Demand for Apollo Lunar Samples," by NASA's Lunar Exploration Analysis Group (LEAG).[1] To create asteroid regolith simulant, and to evaluate it by this Figure of Merit system, it is necessary to have a reference model of asteroid regolith's particle size distribution. We identified four data sets to analyze for the reference model: (1) regolith particles returned by Hayabusa from asteroid Itokawa, which provides a power law, a minimum size, and a maximum size subject to the sampling method; (2) infrared observations of asteroids, which provide mean particle sizes of the surface material as a function of asteroid diameter and class; (3) analyses of observed dust dispersion from disrupted asteroids; and (4) the measurements of grain sizes in regolith breccia meteorites. The only returned sample of asteroid regolith to-date was from the Hayabusa spacecraft at near-Earth asteroid (25143) Itokawa (Tsuchiyama et al., 2011). This sample represents only one type of terrain on one size and class of asteroid in one solar system environment, so we do not adequately understand the range of possible particle size distributions. There is also a question whether the sampling method may have biased the size distribution because it did not function as planned (Yoshikawa et al., 2015). As new data and returned samples become available the reference model we present here should be updated.

---

[1] http://www.lpi.usra.edu/leag/reports/SIM_SATReport2010.pdf. 2010.

**Variations in Asteroid Regolith**

For particle size distributions we found it necessary to distinguish between the surface lag deposit on an asteroid versus regolith material in the bulk. Regolith is created and continuously reworked by the collisional environment of the asteroids (Cheng, 2004). Close encounters with major planets may also reset the structure of a rubble pile asteroid by disrupting and mixing regolith (Richardson, et al., 1998) and possibly releasing subsurface fines to be swept away by the solar wind (Sickafoose, et al., 2002). Following each such disruption, thermal cracking creates new fines at the asteroid's surface (Delbo et al., 2014; Eppes et al., 2010) but only within the penetration depth of the thermal wave, typically a few centimeters, so regolith near the surface should become finer than the regolith beneath it, and this enhancement of fines should be more rapid closer to the surface where the thermal wave's amplitude is greater. The fines may fall into the subsurface via gravity and vibration while restrained by cohesion (Britt and Consolmagno, 2001). Solar wind and photoionization may also winnow these fines off the surface where sunlight penetrates to charge them (Lee, 1996), so the surface may "deflate", developing a lag surface of gravel-sized particles in the sunlight overlying the fines layer, preventing further stripping and deflation. Keihm et al. (2012) using data from Rosetta's flyby of (21) Lutetia found the regolith has lower thermal inertia in the upper few centimeters, increasing with depth "in a manner very similar to that of Earth's Moon," which is consistent with the hypothesis that asteroid surface material is physically different than the bulk.

A list of symbols for the model is provided in Table 1.

**Table 1**
List of Symbols

| Symbol | Description | Units |
| --- | --- | --- |
| $n^S(D)$ | Differential number density for surface regolith | m-1 |
| $n^B(D)$ | Differential number density for bulk regolith | m-1 |
| $n_{\text{Ref}}(D)$ | Reference model differential number density | m-1 |
| $N_C^S(D)$ | Cumulative number density for surface regolith | m-1 |
| $N_C^B(D)$ | Cumulative number density for bulk regolith | m-1 |
| $F^S(D)$ | Differential volume fraction for surface regolith | m-1 |
| $F^B(D)$ | Differential volume fraction for bulk regolith | m-1 |
| $D$ | Particle diameter | m |
| $D_{\text{Min}}^S$ | Minimum particle size of simulant representing surface regolith | m |
| $D_{\text{Max}}^S$ | Maximum particle size of simulant representing surface regolith | m |
| $D_{\text{Min}}^B$ | Minimum particle size of simulant representing bulk regolith | m |
| $D_{\text{Max}}^{\text{B}}$ | Maximum particle size of simulant representing bulk regolith | m |
| $D_{16}$ | Particle diameter at which 16%wt of the regolith is finer | m |
| $D_{50}$ | Median diameter; the diameter at which 50%wt of the regolith is finer | m |
| $D_{84}$ | Particle diameter at which 84%wt of the regolith is finer | m |
| $\widehat{D}$ | Geometric mean particle size | m |
| $\widehat{D}_{\text{V}}$ | Volume-weighted (or mass-weighted) mean particle size | m |
| $\widehat{D}_{\text{S}}$ | Surface-area-weighted mean particle size | m |
| $D_{mn}$ | A type of particle average size, defined in the text | m |
| $\langle D^m \rangle$ | A type of particle average size, defined in the text | m |
| $\delta$ | Asteroid diameter | km |
| $\varepsilon$ | Parameter to scale the surface area occupied by non-round particles | |
| $g$ | Gravity | m/s2 |
| $q$ | Power index | |
| $q^*$ | Effective power index calculated for HED meteorites | |

**Hayabusa Returned Sample and Imagery**

The Hayabusa mission found the surface of Itokawa included smooth terrain that covered about 20% of the surface and rough terrain that covered the rest. Boulders were counted in Hayabusa imagery in which blocks >5 m could be resolved, cobbles were counted in several locations in close-up imagery, and fine particles were measured in a returned sample. Boulder distributions were found to have cumulative power index $-3.1 \pm 0.1$ (Michikami et al., 2008) or $-3.3 \pm 0.1$ (Mazrouei et al., 2014) when measuring the mean horizontal dimension. The distribution becomes shallower for blocks smaller than 6 m, taking a slope of 0 by $D < 2$ m, but this may be an artifact of incomplete counting at the limits of resolution (Mazrouei et al. 2014). The boulder distribution has cumulative power index $-2.8$ when measuring maximum horizontal dimension instead of mean horizontal dimension (Saito et al., 2006), as discussed by Michikami et al. (2008).

Miyamoto et al. (2007) counted cobbles in close-up imagery from Hayabusa and found they fit a power law consistent with cumulative power index $-2.8$ from ~0.2 m to ~2 m. This was based on maximum horizontal dimension and agrees with the slope of Saito et al. (2006), which also was based on maximum horizontal dimension. To combine the boulder and cobble data sets we will use Saito et al. (2006) for consistency in the type of diameter that is used.

The fines sample returned from Itokawa is not well understood because the sampling operation did not function as planned (Yoshikawa et al., 2015) and it is unclear whether electrostatic forces or other factors may have resulted in particle size segregation. After return to Earth, the sample was removed from the sampler by tapping followed by swiping with a spatula, which segregated the particles into coarser and finer fractions (Tsuchiyama et al., 2011). The finer "spatula sample" < 1 µm to ~20 µm was found to have a cumulative power law index of $-2.8$ (differential index $-3.8$), while the coarser "tapping sample" ~10 µm to 100 µm had a cumulative index $-2.0$ (differential $-3.0$) (Tsuchiyama et al., 2011). The change in slope may be an artifact of segregating the particles in Earth's gravity because the change in slopes occurs where gravity separated them. When the spatula and tapping data sets are mathematically recombined as shown in Fig. 1 they collectively follow a cumulative index $-1.5$ (differential

−2.5). The smallest particles identified in the Hayabusa sample are submicron and can potentially be undercounted by cohesively bonding with larger particles. The largest collected particle ~100 µm depended on the collection method so it is possible this power index extends far above 100 µm.

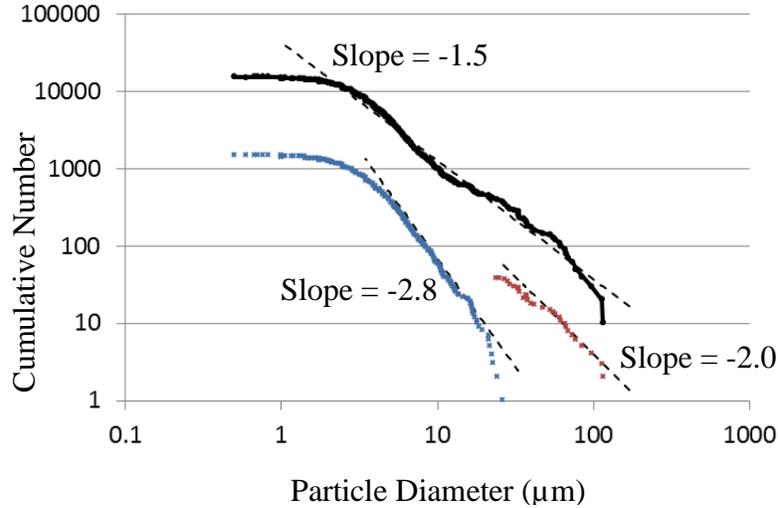

**Figure 1.** Cumulative size distribution of Itokawa particles. Blue (spatula) and red (tapping) and their fitting slopes follow Tsuchiyama et al. (*2011*). Black is these two data sets combined, moved higher ×10 for clarity.

Our goal is to create a math model that spans these three data sets (boulders, cobbles, and fines). The cobbles and boulders share a common power index (Miyamoto et al., 2007). There is a large size gap from the largest returned fine particle at 100 µm and the smallest counted cobble at 0.2 m, and the power index must change somewhere in the gap. The simplest model is to assume only these two power indices exist,

$$n^S(D) = \begin{cases} A_1 \, D^{-2.5}, & 1 \text{ µm} < D \leq D_1 \\ A_2 \, D^{-3.8}, & D_1 < D \geq 50 \text{ m} \end{cases} \quad (1)$$

and the differential distribution is continuous at the transition,

$$A_1 \, D_1^{-2.5} = A_2 \, D_1^{-3.8} \quad (2)$$

The 50 m upper size limit in Eq. (1) is the diameter of the largest boulder observed on Itokawa (Saito et al., 2006). $A_2 = 419{,}000$ is determined by integrating Eq. (1) from 5 m to 50 m and equating it to the observed cumulative value at 5 m in Fig. 2 of Michikami et al. (2008). (The returned fines sample and the cobble size distributions were not exhaustive counts of Itokawa's surface particles so they do not constrain $A_1$ directly.) Since $n^S(D)$ is an extensive count of surface particles, the value of $D_1$ is constrained by the total surface area of Itokawa,

$$\int_{1\,\mu m}^{D_1} A_1\, D^{-2.5} \left(\frac{\varepsilon\pi D^2}{4}\right) dD + \int_{D_1}^{50\,m} A_2\, D^{-3.8} \left(\frac{\varepsilon\pi D^2}{4}\right) dD = 393{,}000 \text{ km}^2 \quad (3)$$

where $\varepsilon$ accounts for the cross-sectional surface area of non-circular particles. Fines in the Itokawa returned sample have average aspect ratio $b/a = 0.71$ (intermediate over longest dimension), and $c/a = 0.43$ (shortest over longest dimension) (Tsuchiyama et al., 2011), but they contribute negligible fraction of the surface area as shown in Fig. 2. Boulders on Itokawa in the 0.1 to 5 m range have average $b/a = 0.68$, and in the range >5 m have $b/a = 0.62$ (Michikami et al., 2010), which correspond to the values of $\varepsilon$ if the particles are treated as ellipsoids lying "flat" ($c$ direction up). We use $\varepsilon = 0.67$ in rough approximation for all sizes, which yields $D_1 = 0.470$ m and $A_1 = 338{,}700$ from Eq. (2) and Eq. (3). Integrating Eq. (1) produces the cumulative distribution

$$N_C^S(D) = \begin{cases} 225{,}800\, D^{-1.5} - 325{,}600, & 1\,\mu m < D \leq 0.470\,m \\ 44{,}298.7\, D^{-2.8}, & 0.470\,m < D \leq 50\,m \end{cases} \quad (4)$$

This is plotted in Fig. 2 compared against the data for fines (Tsuchiyama et al., 2011), cobbles (Miyamoto et al., 2007), and boulders (Saito et al., 2006). The abrupt slope change of the differential distribution at $D_1$ produces in the cumulative distribution a curve through the range 0.2 m to 0.5 m, which fits the cobble distribution arguably better than a straight power law, as shown in Fig. 3. The $D_1$ had to be located inside the range of the cobble distribution in order to keep the net surface area from exceeding the surface area of Itokawa. That eliminates the possibility that another power index intermediate to the 2.5 and 3.8 indices could exist within the large gap between fines and cobbles, providing some confidence in the two-slope model. The excellent fit across all data sets also provides some confidence in the continuity assumption.

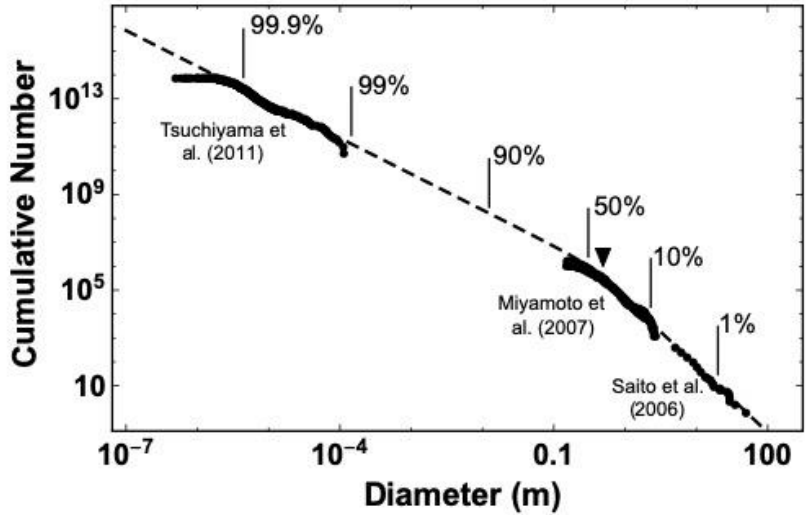

**Figure 2.** Cumulative particle size distribution. Dashed: Model per Eq. (4). Filled circles: three data sets from Itokawa. Annotations show percent coverage of Itokawa's surface by particle sizes finer-than, per integration of the model. Triangle marks $D_1$, the location of the change in slope in the model's differential distribution.

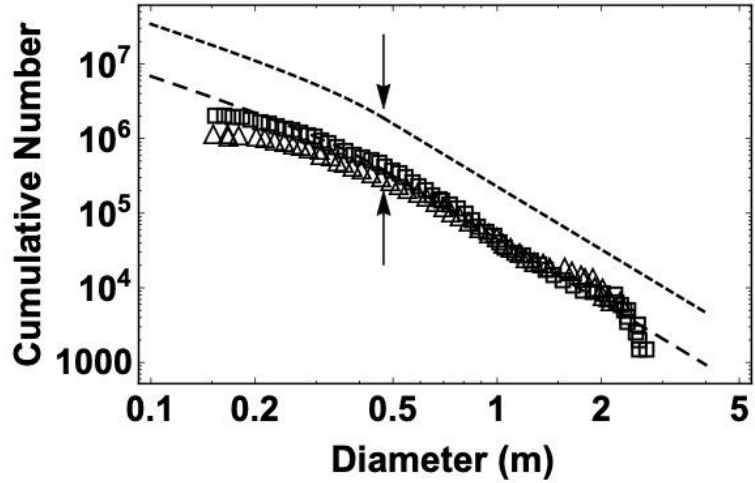

**Figure 3.** Cumulative size distribution focusing on the region of cobbles. Long dashed: Model per Eq. (4). Short dashed: the model raised a factor of 5 for clarity to show the curvature. Squares and triangles: both data sets per Miyamoto et al. (2007), moved vertically to best match the model since their amplitude is otherwise not determined. Arrows: location of the change in slope in the model's differential distribution. The gentle curve resulting in the cumulative distribution matches the data well.

The volume-weighted particle size distribution function for the surface deposit is,

$$F^S(D) = \frac{n^S(D)\,D^3}{\int_{\sim 0}^{50\text{ m}} n^S(D)\,D^3 dD} = \begin{cases} 0.37047\, D^{+0.5}, & 1\ \mu\text{m} < D \leq 0.470\text{ mm} \\ 0.13873\, D^{-0.8}, & 0.470\text{ mm} < D \leq 50\text{ m} \end{cases} \quad (5)$$

Integrating finds the $D_X$ parameters of geotechnical engineering, defined as the particle diameters such that X% of the regolith's mass resides in smaller particles,

$$\int_{\sim 0}^{D_X} F^S(D)\,dD = X\% \quad (6)$$

We find $D_{16} = 0.885$ m, $D_{50} = 6.77$ m, and $D_{84} = 28.6$ m. The *median particle size* in geotechnical engineering is $D_{50}$, while the *geometric mean* is $\widehat{D} = \sqrt[3]{D_{16} D_{50} D_{84}} = 5.56$ m. We also calculate the volume-weighted average particle size,

$$\widehat{D}_V = \frac{\int_{\sim 0}^{50\text{ m}} n^S(D) D^4\, dD}{\int_{\sim 0}^{50\text{ m}} n^S(D) D^3\, dD} \quad (7)$$

and the surface-area-weighted average particle size,

$$\widehat{D}_S = \frac{\int_{\sim 0}^{50\text{ m}} n^S(D)D^3\, dD}{\int_{\sim 0}^{50\text{ m}} n^S(D)D^2\, dD} \tag{8}$$

finding $\widehat{D}_V = 12.6$ m and $\widehat{D}_S = 1.22$ m.

**Infrared Observations**
For the infrared observations of asteroid regolith, the theory of Gundlach and Blum (2013), hereafter GB, attempts a correlation between mean particle size and observations of globally averaged thermal inertia. The results are replicated in Fig. 4, with an added fitting function for the smaller asteroids and another for the larger asteroids,

$$\widehat{D} = \begin{cases} 1277\, g^{-0.32}, & g \leq 0.0185 \text{ m/s}^2 \\ 4.43\, g^{-1.74}, & g \geq 0.0185 \text{ m/s}^2 \end{cases} \text{(\textmu m)} \tag{9}$$

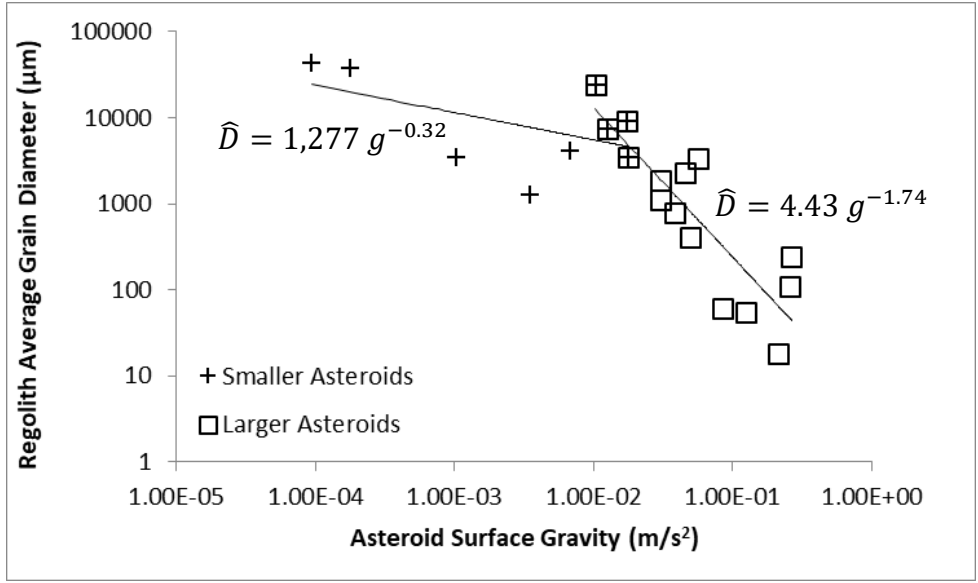

**Figure 4.** Regolith coarseness varying with asteroid size. Data reproduced from Gundlach and Blum (2013) with added scaling laws fitted to small and large asteroids. Four asteroids in the overlap region were included in both small and large populations.

Ideally, a theory like this could be used for the asteroid simulant reference model. However, we discovered there is a problem reconciling GB with the size distribution in (1) or similar equation that includes boulders normalized to the surface area of the asteroid. The resulting mean diameters from Eqs. (5) – (8) are two orders of magnitude higher than the $4.2^{+0.6}_{-2.8}$ cm mean diameter predicted by GB (Gundlach and Blum, 2013). This raises the question what globally averaged "mean size" means. There are many definitions of mean size depending on how the particle size appears in the equations and how the equations are averaged. The equations in GB use particle size with exponent unity in the radiation term and inverse unity in the conduction term, so presumably an average over each term would result in two different mean values, $\langle D \rangle$ and $\langle D^{-1}\rangle^{-1}$, where the angle brackets represent a type of averaging. This could be the volume-weighted average as in $\widehat{D}_V$, or the area-weighted average as in $\widehat{D}_S$, or perhaps the geotechnical median or geometric mean, both based on mass fractions (Carrier et al., 1991) or any mean diameter of the form

$$D_{mn} = \left(\frac{\langle D^m\rangle}{\langle D^n\rangle}\right)^{1/(m-n)} \tag{10}$$

for all *m* and *n*, where the *m*th moment of the differential size distribution is

$$\langle D^m\rangle = \int_{D_{\min}}^{D_{\max}} n(D)\, D^m \mathrm{d}D \tag{11}$$

We have tried all these but found no definition of mean value that works for GB predictions within an order of magnitude for both the Itokawa distribution in Eq. (1) and the lunar data set that was used to validate GB.

A second problem is that the lunar soil measured in laboratories to validate GB had been pre-sieved to remove larger particles (McKay et al. 1974; McKay et al. 2009) so its sizes ranged only ~1 μm to 1 mm (3 orders of magnitude), much narrower than Itokawa's in situ particles ~1 μm to ~50 m (7.5 orders of magnitude). Cobbles and blocks have much larger diameter than the depth of the thermal wave into the subsurface fine particles (approximately 5 cm based on our modeling of Itokawa), so it is problematic to average embedded boulders and cobbles >5 cm together with fines for a mean size in thermal conductivity equations. Particles >5 cm cover 79.8% of the asteroid's surface according to Eq. (4). Until additional theoretical progress is made, we do not believe we can use globally averaged thermal observations to extrapolate Eq. (9) from Itokawa to other sizes of asteroids, although GB shows there is a size dependence.

**Disrupted Asteroids**

For the size distribution of bulk regolith below the weathered surface, we rely on analyses of the active asteroids. There are multiple mechanisms that may eject dust from asteroids, including sublimation of ice, thermal release of water from hydrated minerals, and electrostatic forces. Jewitt (2012) provides an extensive review. We focus on four cases that appear to have been caused by collisions or rotational disruption: P/2010 A2 (LINEAR), P/2012 F5 (Gibbs), 596 Scheila, and a zodiacal dust band possibly associated with the Emilkowalski cluster. For example, P/2010 A2 (LINEAR) has been interpreted as an asteroid that was impulsively disrupted producing a long tail of ejecta (Jewitt et al., 2010; Snodgrass et al., 2010; Kleyna et al., 2013), or possibly having spun up under YORP it reached its centripetal limit and so was rotationally disrupted (Jewitt, 2012). Moreno et al. (2010) estimated the asteroid's diameter is $\delta = 0.2 - 0.3$ km and the ejecta mass is 4% to 20% of the non-ejected mass. Therefore, if ejection is assumed uniform across the sphere (worst case for sampling bulk material), the outermost 2.6 to 17.7 m of material was ejected, and the vast majority of the material would be well below the weathered layer modified by thermal cracking and deflation. Assuming this thermal wave weathered layer is 5 cm thick, the weathered material would constitute only 0.3 - 2% of the ejecta mass, not a significant contributor. If ejection is from a localized crater, then the surficial contribution is far less. Therefore, disrupted asteroid tails can be interpreted as samples of bulk material.

For P/2010 A2 (LINEAR), by comparing numerical simulations to observations of the particles in the tail dispersing via radiation pressure, the ejected particle diameters have been found to range from $D_{min} = 0.6$ to $D_{max} = 40$ mm (Hainaut et al., 2012; Jewitt et al., 2013). (We converted grain radii to grain diameter, following convention of the simulants literature.) The modeling shows the differential particle size distribution is consistent with a power law having index $q = -3.3 \pm 0.2$ (Jewitt et al., 2010), $q = -3.4$ (Snodgrass et al., 2010), $q = -3.4 \pm 0.3$ (Moreno et al, 2010), $q = -3.44 \pm 0.08$ (Hainaut et al., 2012), or $q = -3.5 \pm 0.1$ (Jewitt et al., 2013). A weighted average of the four of these that provide uncertainties, where the weight factor is the inverse of the uncertainty squared (and the net uncertainty is the square root of the inverse of the sum of the weights), results in $q = -3.45 \pm 0.06$, which we take as the best estimate.

For object P/2012 F5 (Gibbs), Stevenson et al. (2012) found $\delta = \sim 1 - 2.9$ km and ejecta $D_{min} = 40$ µm although smaller particles may have been below the detection threshold, with $D_{max} >$ cm-sized. Moreno et al. (2012) found $q = -3.7 \pm 0.1$, $D_{min} = 130 - 180$ µm (smaller may have existed below the detection threshold), and $D_{max} = 56 \pm 20$ cm. We take $D_{min} = 40$ µm, $D_{max} = 56$ cm, and $q = -3.7$ as the best estimates.

For object 596 Sheila, Moreno et al. (2011) found $\delta = 113$ km with ejecta $D_{min} = 1.6$ µm, $D_{max} = 10$ cm, and $q = -3$ (no uncertainty given). Ishiguro et al. (2011) found that $D_{max} = 280$ µm, and $q = -3.5$ produced models that fit the observations, but 3.0 and 4.0 did not fit the models, so we state their estimate as $q = -3.5 \pm 0.5$ and we adopt it as the best estimate. The estimates of $D_{max}$ disagree by 2.5 orders of magnitude. Moreno et al. (2011) show that the smaller value cannot work in their model, but Ishiguro et al. (2011) do not give details how their model produced $D_{max}$ so we cannot evaluate. Therefore, we provisionally use $D_{max} = 10$ cm and $D_{min} = 1.6$ µm.

For the zodiacal band, Kehoe et al. (2015) identified the likely source as the parent body of the Emilkowalski cluster, $\delta \sim 10$ km, in a collision that occurred significantly less than 1 million years ago. Modeling finds differential size distribution $q = -3.1$ (cumulative index -2.1). This is not as steep and is therefore coarser than the size distributions of the other disrupted asteroids. Kehoe et al. argue the fines are being removed by radiation pressure, which has occurred over a significantly longer time compared to the other three disrupted asteroids. These findings are summarized in Table 2.

**Table 2. Best Estimates of Disrupted Asteroid Parameters.**

| Body | Nucleus Diameter (km) | $D_{min}$ (mm) | $D_{max}$ (mm) | $q$ | $D_{50}$ (mm) Note 1 | $\widehat{D}_V$ (mm) Note 2 | $\widehat{D}_S$ (mm) Note 3 |
|---|---|---|---|---|---|---|---|
| P/2010 A2 (LINEAR) | 0.2 – 0.3 | 0.6 | 40 | -3.45 | 13.5 | 15.7 | 5.2 |
| P/2012 F5 (Gibbs) | ~ 1 – 2.9 | 0.04 | 560 | -3.7 | 66.8 | 137 | 1.5 |
| 596 Scheila | 113 | 0.0016 | 100 | -3.5 | 25.2 | 33.5 | 0.40 |
| Emilkowalski Cluster | ~ 10 | -- | -- | -3.1 | -- | -- | -- |
| Itokawa (Hayabusa) | | | | | 6,770 | 12,600 | 1,220 |

Note 1: The particle diameter for which 50% of the mass is finer, calculated by integration using $D_{min}$, $D_{max}$, and $q$ from the table.
Note 2: The volume-averaged mean particle diameter per Eq. (7), calculated by integration using $D_{min}$, $D_{max}$, and $q$ from the table.
Note 3: The particle surface area-averaged mean particle diameter per Eq. (8), calculated by integration using $D_{min}$, $D_{max}$, and $q$ from the table.

We integrated the differential size distribution functions defined by Table 2 to find several versions of median or average particle size, also reported in Table 2. In each case the values are one or more orders of magnitude finer than the corresponding values calculated for Itokawa's surface material. This is probably because $D_{min}$ and $D_{max}$ measured from disrupted asteroids depended upon telescope detection limits and ejecta escape velocities, so they probably do not represent particle size limits of the bulk regolith. It is possible $D_{min}$ and $D_{max}$ in bulk regolith vary with size of the asteroid as suggested for the surficial deposit by GB, but data are inadequate to constrain them. The power law indices for disrupted asteroids are very close to the value -3.5 predicted by Dohnanyi (1969) for collisional equilibrium and is remarkably constant across size ranges of the parent bodies, suggesting that bulk asteroid regolith is in fact in collisional equilibrium despite a non-equilibrium lag deposit at the surface.

**Regolith Breccia Meteorites**

Finally, we examine the brecciated HED meteorites, which as regolith breccias are fossilized regolith. It is believed the parent body of most HEDs is Vesta (McSween et al., 2010), which is not the typical gravity environment for smaller asteroids. Several studies have analyzed HED texture (Labotka and Papike, 1980; Fuhrman and Papike, 1982; Pun et al., 1998) and counted particles into three bins: $N_1$, 20 μm to 200 μm; $N_2$, 200 μm to 2000 μm; and $N_3$, >2000. Some also reported volume of material < 20 μm but without particle counts. This coarse binning is inadequate to calculate a particle size distribution, but if we assume they have a power law that extends over the middle two bins the effective index can be calculated,

$$q^* = \log_{10} \frac{N_2}{N_1} - 1 \qquad (12)$$

Table 3 shows the available data and the calculated values of $q^*$. The mean value of the last column is $\overline{q^*} = -0.82 \pm 0.28$, so whether or not it is actually a power law we conclude the parent body's regolith is very far from collisional equilibrium and therefore dissimilar to the bulk regolith of the smaller, disrupted asteroids.

**Table 3.** Particle Sizes in HED Meteorites.

| Meteorite | Reference | $N_1$ | $N_2$ | $N_3$ | Vol% <20 μm | $q^*$ |
|---|---|---|---|---|---|---|
| Yurtuk | Labotka and Papike, 1980. | 378 | 200 | 11 | 44.2%* | -1.28 |
| Frankfort | Labotka and Papike, 1980. | 393 | 806 | 4 | 29.2%* | -0.69 |
| ALHA 77302 | Labotka and Papike, 1980. | 462 | 614 | 1 | 38.2%* | -0.88 |
| Pavloka | Labotka and Papike, 1980. | 265 | 517 | 2 | 39.6%* | -0.71 |
| Malvern | Labotka and Papike, 1980. | 400 | 331 | 12 | 21.1%* | -1.08 |
| Kapoeta | Fuhrman and Papike, 1982 | 506 | 1574 | 18 | 26.8% | -0.50 |
| Bununu | Fuhrman and Papike, 1982 | 614 | 342 | 5 | 25.3% | -1.25 |
| Bholgati | Fuhrman and Papike, 1982 | 493 | 1411 | 8 | 24.4% | -0.54 |
| ALHA 76005 | Fuhrman and Papike, 1982 | 343 | 500 | 2 | 28.5% | -0.84 |
| Kapoeta | Pun et al., 1998. | 2472 | 9813 | 13517 | 41.7% | -0.40 |
| Kapoeta Clast A (mafic breccia) | Pun et al., 1998. | 305 | 424 | 180 | 52.2% | -0.85 |
| Kapoeta Clast D (howardite) | Pun et al., 1998. | 496 | 818 | 268 | 54.1% | -0.78 |

* Calculated from the reported data

For comparison, the lunar regolith is another surface that is not in collisional equilibrium. Fitting a power law to the average size distribution of lunar soil from Carrier (2003), after converting it to differential number of particles, we find its index is $q = -4.8$ for $D > 100$ μm. This indicates lunar soil is overly fine compared to collisional equilibrium, whereas the HED textures are overly coarse. Perhaps the HED regolith was very immature and buried by subsequent collisions. In any case, the data disagree with the power law from disrupted asteroids so we choose not to base our reference model on the HEDs.

**Conclusion**

Based on these considerations, we define "Version 1.0" of an asteroid regolith simulant reference model,

$$n_{\text{Ref}}(D) = \begin{cases} (1/c_1)\, D^{-2.5}, & D_{\text{Min}}^{\text{S}} \geq D \geq D_{\text{Max}}^{\text{S}}, \quad \text{for surface deposits} \\ (1/c_2)\, D^{-3.5}, & D_{\text{Min}}^{\text{B}} \geq D \geq D_{\text{Max}}^{\text{B}}, \quad \text{for bulk regolith} \end{cases}$$

$$c_1 = \int_{D_{\text{Min}}^{\text{S}}}^{D_{\text{Max}}^{\text{S}}} D^{-2.5} dD, \quad c_2 = \int_{D_{\text{Min}}^{\text{B}}}^{D_{\text{Max}}^{\text{B}}} D^{-3.5} dD \tag{13}$$

where $D_{\text{Min}}^{\text{S}}$, $D_{\text{Max}}^{\text{S}}$, $D_{\text{Min}}^{\text{B}}$, and $D_{\text{Max}}^{\text{B}}$ depend upon the asteroid size and its history, but they are unconstrained until more data are available and they are treated as user-selectable parameters for simulant users. This is a very simplistic model that neglects the differences in smooth versus blocky terrains of an asteroid, for example, but it is a starting point that will be iterated and improved as more data become available. An asteroid regolith simulant can be scored per

NASA's Figure of Merit system by how closely it matches these power laws for replicating either surficial or bulk regolith. Simulant users can modify an asteroid simulant to match a desired asteroid's $D_{min}$ and $D_{max}$ according to their model of the target body by sieving to reduce the simulant's particle size range or by crushing cobbles/regolith as additives to extend its particle size range, if desired. Metzger et al. (2019) present the particle size distribution of a recently developed asteroid regolith simulant and they show how this reference model was used in calculating the Figure of Merit for that simulant.

**Acknowledgement**

This work was directly supported by NASA's Solar System Exploration Research Virtual Institute cooperative agreement award NNA14AB05A. A portion of the work was supported by NASA's Small Business Innovative Research (SBIR) program contract number NNX15CK10P, "Task-Specific Asteroid Simulants for Ground Testing." Declaration of interest: none.

**References**

1. Britt, D. T., and G. J. Consolmagno. "Asteroid bulk density: Implications for the structure of asteroids." Abstract no. 1611. 32nd Annual Lunar and Planetary Science Conference, March 12-16, 2001, Houston, Texas.

2. Carrier III, W. David. "Particle size distribution of lunar soil." *Journal of Geotechnical and Geoenvironmental Engineering* 129, no. 10 (2003): 956-959.

3. Carrier, W. David, III, Gary R. Olhoeft, and Wendell Mendell. "Physical properties of the lunar surface." *Lunar Sourcebook* (1991): 475-594.

4. Cheng, Andrew F. "Collisional evolution of the asteroid belt." *Icarus* 169, no. 2 (2004): 357-372.

5. Delbo M., Libourel G., Wilkerson J., Murdoch N., Michel P., Ramesh K.T., Ganino C.,Verati C. and Marchi S. (2014) Thermal fatigue as the origin of regolith on small asteroids. *Nature* 50, 233.

6. Dohnanyi, J. S. "Collisional model of asteroids and their debris." *Journal of Geophysical Research* 74, no. 10 (1969): 2531-2554.

7. Eppes, M.C., McFadden, L.D., Wegmann, K.W. and Scuderi, L.A. (2010). Cracks in desert pavement rocks: Further insights into mechanical weathering by directional insolation. *Geomorphology* 123(1), pp.97-108.

8. Fuhrman, Mt, and J. J. Papike. "Howardites and polymict eucrites: Regolith samples from the eucrite parent body. Petrology of Bholgati, Bununu, Kapoeta, and ALHA 76005." In *Lunar and Planetary Science Conference Proceedings*, vol. 12, pp. 1257-1279. 1982.

9. Gundlach, Bastian, and Jürgen Blum. "A new method to determine the grain size of planetary regolith." *Icarus* 223.1 (2013): 479-492.

10. Hainaut, O. R., J. Kleyna, G. Sarid, B. Hermalyn, A. Zenn, K. J. Meech, P. H. Schultz et al. "P/2010 A2 LINEAR-I. An impact in the asteroid main belt." *Astronomy & Astrophysics* 537 (2012): A69.

11. Ishiguro, Masateru, Hidekazu Hanayama, Sunao Hasegawa, Yuki Sarugaku, Jun-ichi Watanabe, Hideaki Fujiwara, Hiroshi Terada et al. "Interpretation of (596) Scheila's Triple Dust Tails." *The Astrophysical Journal Letters* 741, no. 1 (2011): L24.

12. Jewitt, David, Masateru Ishiguro, and Jessica Agarwal. "Large particles in active asteroid P/2010 A2." *The Astrophysical Journal Letters* 764, no. 1 (2013): L5.

13. Jewitt, David, Harold Weaver, Jessica Agarwal, Max Mutchler, and Michal Drahus. "A recent disruption of the main-belt asteroid P/2010 A2." *Nature* 467, no. 7317 (2010): 817-819.

14. Kehoe, AJ Espy, et al. (2015), "Signatures of recent asteroid disruptions in the formation and evolution of solar system dust bands." *The Astrophysical Journal* 811.1: 66.

15. Keihm, S., F. Tosi, L. Kamp, F. Capaccioni, S. Gulkis, D. Grassi, M. Hofstadter et al. (2012). "Interpretation of combined infrared, submillimeter, and millimeter thermal flux data obtained during the Rosetta fly-by of Asteroid (21) Lutetia." *Icarus* 221, no. 1: 395-404.

16. Kleyna, Jan, O. R. Hainaut, and K. J. Meech. "P/2010 A2 LINEAR-II. Dynamical dust modelling." Astronomy & Astrophysics 549 (2013): A13.

17. Labotka, T. C., and J. J. Papike. "Howardites-Samples of the regolith of the eucrite parent-body: Petrology of Frankfort, Pavlovka, Yurtuk, Malvern, and ALHA 77302." In *Lunar and Planetary Science Conference Proceedings*, vol. 11, pp. 1103-1130. 1980.

18. Lee, P. (1996). "Dust levitation on asteroids." *Icarus*, 124(1), pp.181-194.

19. Mazrouei, S., M. G. Daly, Olivier S. Barnouin, C. M. Ernst, and I. DeSouza. "Block distributions on Itokawa." *Icarus* 229 (2014): 181-189.

20. McKay, D. S., B. L. Cooper, and L. M. Riofrio. "New measurements of the particle size distribution of Apollo 11 lunar soil." In *Lunar and Planetary Science Conference*, vol. 40. 2009.

21. McKay, D. S., R. M. Fruland, and G. H. Heiken. "Grain size and the evolution of lunar soils." In *Lunar and Planetary Science Conference Proceedings*, vol. 5, pp. 887-906. 1974.

22. McSween Jr, Harry Y., David W. Mittlefehldt, Andrew W. Beck, Rhiannon G. Mayne, and Timothy J. McCoy. "HED meteorites and their relationship to the geology of Vesta and the Dawn mission." In *The Dawn Mission to Minor Planets 4 Vesta and 1 Ceres*, pp. 141-174. Springer New York, 2010.

23. Metzger, Philip T., Daniel T. Britt, Stephen D. Covey, and John S. Lewis. "Results of the 2015 Workshop on Asteroid Simulants." *Earth and Space 2016: Engineering, Science, Construction, and Operations in Challenging Environments* (2016): 94.

24. Metzger, Philip T., Daniel T. Britt, Stephen Covey, Cody Schultz, Kevin M. Cannon, Kevin D. Grossman, James G. Mantovani, and Robert P. Mueller, "Measuring the Fidelity of Asteroid Regolith and Cobble Simulants," *Icarus* 321 (2019): 632-646.

25. Michikami, Tatsuhiro, Akiko M. Nakamura, Naru Hirata, Robert W. Gaskell, Ryosuke Nakamura, Takayuki Honda, Chikatoshi Honda et al. "Size-frequency statistics of boulders on global surface of asteroid 25143 Itokawa." *Earth, planets and space* 60, no. 1 (2008): 13-20.

26. Michikami, Tatsuhiro, Akiko M. Nakamura, and Naru Hirata. "The shape distribution of boulders on Asteroid 25143 Itokawa: Comparison with fragments from impact experiments." *Icarus* 207, no. 1 (2010): 277-284.

27. Miyamoto, Hideaki, Hajime Yano, Daniel J. Scheeres, Shinsuke Abe, Olivier Barnouin-Jha, Andrew F. Cheng, Hirohide Demura et al. "Regolith migration and sorting on asteroid Itokawa." *Science* 316, no. 5827 (2007): 1011-1014.

28. Moreno, Fernando, Javier Licandro, and Antonio Cabrera-Lavers. "A short-duration event as the cause of dust ejection from main-belt comet P/2012 F5 (Gibbs)." *The Astrophysical Journal Letters* 761, no. 1 (2012): L12.

29. Moreno, F., J. Licandro, J. L. Ortiz, L. M. Lara, V. Alí-Lagoa, O. Vaduvescu, N. Morales, A. Molina, and Z-Y. Lin. "(596) Scheila in outburst: a probable collision event in the Main Asteroid Belt." *The Astrophysical Journal* 738, no. 2 (2011): 130.

30. Moreno, Fernando, Javier Licandro, G-P. Tozzi, José Luis Ortiz, Antonia Cabrera-Lavers, Thomas Augusteijn, Tiina Liimets et al. "Water-ice-driven Activity on Main-Belt Comet P/2010 A2 (LINEAR)?" *The Astrophysical Journal Letters* 718, no. 2 (2010): L132.

31. Pun, Aurora, Klaus Keil, G. Taylor, and Rainer Wieler. "The Kapoeta howardite: Implications for the regolith evolution of the howardite-eucrite-diogenite parent body." *Meteoritics & Planetary Science* 33, no. 4 (1998): 835-851.

32. Richardson, D.C., Bottke, W.F. and Love, S.G. (1998). Tidal distortion and disruption of Earth-crossing asteroids. *Icarus*, *134*(1), pp.47-76.

33. Saito, J., H. Miyamoto, R. Nakamura, M. Ishiguro, T. Michikami, A. M. Nakamura, H. Demura et al. "Detailed images of asteroid 25143 Itokawa from Hayabusa." *Science* 312, no. 5778 (2006): 1341-1344.

34. Schrader, Christian M., Douglas L. Rickman, Carole A. Mclemore, John C. Fikes, Douglas B. Stoeser, Susan J. Wentworth, and David S. McKay. "Lunar regolith characterization for simulant design and evaluation using figure of merit algorithms." In *Proc., 47th AIAA Aerospace Sciences Meeting. ISO*, vol. 690. 2009.

35. Sickafoose, A.A., Colwell, J.E., Horányi, M. and Robertson, S. (2002). Experimental levitation of dust grains in a plasma sheath. *J Geophys Res: Space Phys*, *107*(A11).

36. Snodgrass, C., Tubiana, C., Vincent, J.B., Sierks, H., Hviid, S., Moissi, R., Boehnhardt, H., Barbieri, C., Koschny, D., Lamy, P. and Rickman, H., 2010. A collision in 2009 as the origin of the debris trail of asteroid P/2010A2. *Nature*, *467*, pp.814-816.

37. Stevenson, R., E. A. Kramer, J. M. Bauer, J. R. Masiero, and A. K. Mainzer, "Characterization of active main belt object P/2012 F5 (Gibbs): A possible impacted asteroid." *The Astrophysical Journal* 759 no. 2 (2012): 142.

38. Taylor, Lawrence A., and Yang Liu. "Important considerations for lunar soil simulants." *Earth and Space 2010: Engineering, Science, Construction, and Operations in Challenging Environments* (2010).

39. Taylor L.A., Pieters C.M., and Britt D.T. (2016) Evaluations of lunar regolith simulants *Planetary and Space Science*, Volume 126, July 2016, Pages 1–7.

40. Tsuchiyama, Akira, et al. (2011). "Three-dimensional structure of Hayabusa samples: origin and evolution of Itokawa regolith." Science 333.6046: 1125-1128.